\documentclass{emulateapj}

\usepackage{graphics,epsf,graphicx,longtable,natbib,times,mathptm}

\newcommand{\kms}{\ifmmode{~{\rm km\,s}^{-1}}\else{~km~s$^{-1}$~}\fi}
\newcommand{\msun}{\ifmmode{{\rm M}_\odot}\else{${\rm M}_\odot$}\fi}

\slugcomment{to appear in The Astrophysical Journal}

\shorttitle{1D hydrodynamical models of wind driven shocks}
\shortauthors{Vaytet et al.}

\begin{document}

\title{Swift observations of the 2006 outburst of the recurrent nova RS Ophiuchi:\\
       II. 1D hydrodynamical models of wind driven shocks}

\author{N. M. H. Vaytet, T. J. O'Brien}
\affil{Jodrell Bank Observatory, School of Physics and Astronomy, The University of Manchester, Macclesfield, Cheshire, SK11 9DL, UK}
\email{neil.m.vaytet@postgrad.manchester.ac.uk}
\email{tim.obrien@manchester.ac.uk}

\and

\author{M. F. Bode}
\affil{Astrophysics Research Institute, Liverpool John Moores University, Twelve Quays House, Egerton Wharf, Birkenhead, CH41 1LD, UK\\}
\email{mfb@astro.livjm.ac.uk}

\begin{abstract}
Following the early {\it Swift} X-ray observations of the latest outburst of the recurrent nova RS Ophiuchi in February 2006 (Paper I), we present new 1D hydrodynamical models of the system which take into account all three phases of the remnant evolution. The models suggest a novel way of modelling the system by treating the outburst as a sudden increase then decrease in wind mass-loss rate and velocity. The differences between this wind model and previous Primakoff-type simulations are described. A more complex structure, even in 1D, is revealed through the presence of both forward and reverse shocks, with a separating contact discontinuity. The effects of radiative cooling are investigated and key outburst parameters such as mass-loss rate, ejecta velocity and mass are varied. The shock velocities as a function of time are compared to the ones derived in Paper I. We show how the manner in which the matter is ejected controls the evolution of the shock and that for a well-cooled remnant, the shock deceleration rate depends on the amount of energy that is radiated away.
\end{abstract}


\section{Introduction}\label{s1}

RS Ophiuchi is part of the small class of Recurrent Novae (RN) with only 10 known members \citep{anupama02}. The central system is a binary comprising a white dwarf (WD) and a red giant (RG) companion \citep{dobrzycka94}. The generally accepted model for nova outbursts \citep{rose67} involves mass transfer from the companion to the WD. The build-up of pressure and temperature in the degenerate layer of accreted hydrogen eventually leads to a thermonuclear runaway (TNR) on the surface of the WD, resulting in the high-speed ejection of a shell of material into the circumstellar medium \citep{starrfield89}. The shock interaction of the ejecta with the surrounding medium has been found to heat the gas to temperatures of $10^{7} - 10^{8}$K, yielding hard X-ray radiation (\citealt{lloyd92}; \citealt{o'brien94}; \citealt{mukai01}). In the case of RS Ophiuchi, the ejecta run into the surrounding dense RG wind. Soft X-ray emission is also expected to be revealed later in the outburst from a central WD close to Eddington luminosity (\citealt{krautter96}; \citealt{balman98}).

RS Ophiuchi has undergone recorded outbursts in 1898, 1933, 1958, 1967, 1985 (see \citealt{rosino87a}; \citealt{rosino87b}) and most recently on February 12 2006 \citep{hirosawa06}, with possible additional outbursts in 1907 and 1945 (\citealt{schaeffer04}; \citealt{oppenheimer93}). Its binary system has an orbital period of $455.72\pm 0.83$ days \citep{fekel00}. The mass of the WD has been measured to be close to the Chandrasekhar limit by Osborne et al. (in prep.) who found $M_{WD} \simeq 1.4~\msun$ from Xray observations, and \citet{hachisu06} who obtained $M_{WD} = 1.35~\msun$ from a detailed optical lightcurve analysis. It lies at a distance of $1.6\pm0.3$ kpc, derived from several methods \citep{bode87}.

Observations of RS Ophiuchi prior to 1985 were purely optical and it had been predicted that observations at other wavelengths would present much new information. In 1985, RS Ophiuchi was observed across the electromagnetic spectrum, from radio to X-rays. Radio observations started 18 days after outburst, detecting the source at a suprisingly high flux of 23 mJy. The radio lightcurve at 4.9 GHz was found to peak 37 days after outburst (at $\sim 60$~mJy) and decay to half power after a further 40 days (\citealt{davis87}; \citealt{hjellming86}). EXOSAT X-ray observations covered the period $55 - 251$ days after outburst \citep{mason87} and revealed a very rapidly evolving behaviour. Analytical spherically symmetric models of \citet{bode85} based on the X-ray and early radio observations led to analogies between RS Ophiuchi and young supernova remnants, although the RN was found to evolve on much shorter timescales. The models suggested the presence of several regions with different temperatures, consistent with the infrared observations of \citet{evans88}. Supernova-type analytical models including the effect of radiative heat-loss were derived by \citet{o'brien87}, who predicted that the shock wave would reach the edge of the RG wind some 65 days after outburst, seemingly consistent with the optical spectroscopic observations of \citet{anupama89} who suggest the shock overtook the RG wind between 60 and 90 days after outburst. Numerical models of \citet{o'brien92} yielded estimates for key parameters of the RS Ophiuchi system such as the outburst energy $E_{0} = 1.1\times 10^{43}$erg and the ejected mass $M_{\mathrm{ej}} = 1.1\times 10^{-6} \msun$. Their studies concluded that the remnant is expected to evolve rapidly ($\sim 6$ days) through the phase of free expansion (Phase I), and that day 55 was situated during the transition between Phases II (hot adiabatic blast wave) and III (cooled remnant). \citet{o'brien92} also investigated the explanation of \citet{mason87} for a sharp decrease in X-ray flux around day 70 after outburst by modelling the shock breaking out of the RG wind. \citet{contini95} studied the late behaviour of RS Ophiuchi around 200 days after outburst, and their models showed that the late X-ray flux at this time could be accounted for by shock emission only. They also predicted a high He abundance in the shell of the remnant.

On February 12.83UT 2006, RS Ophiuchi was observed to reach a magnitude $V=4.5$ \citep{hirosawa06}. This time, X-ray observations were quickly underway starting just 2 days after outburst with the RXTE satellite \citep{sokoloski06} and 3.17 days after outburst using the XRT instrument on board the {\it Swift} telescope (\citealt{bode06}, hereafter Paper I). Further analysis of the {\it Swift} data showed that the outburst itself had been captured in the two lowest energy channels of the hard X-ray Burst Alert Telescope. This gave an insight into the very early stages of the remnant evolution which were previously unobserved. RS Ophiuchi was seen as an initially bright source of hard X-rays, gradually softening with time. Initial analysis suggests that the basic shock model for the early X-ray emission is correct with Phase I terminating at $\sim 6$ days, but with a very rapid transition from Phase II to Phase III thereafter. Around 26 days after outburst, a totally new soft component appeared in the spectrum which was previously unaccounted for. In 1985, as X-ray observations only started 55 days after outburst, this extra source of X-ray flux would have already been present as opposed to being observed to appear some time after outburst. It was most probably detected but all of the emission was attributed to the shocks. Due to its late emergence and soft spectrum radically different to the shocks' hard X-ray spectra, it is most likely that the emission has a different origin. This component has been attributed to the WD undergoing a Super Soft Source (SSS) phase of nuclear burning (\citealt{osborne06a}; \citealt{hachisu07}).

In an effort to improve on the models of the 1985 outbursts and, in particular, to address the new observations of Phase I, we present in this paper revised hydrodynamical models for RS Ophiuchi where the outburst results in mass-loss in the form of a fast wind which runs into the surrounding slow RG wind. This scheme takes into account the ejection of material in the outburst as well as allowing the duration of the fast wind phase to be varied, as opposed to the instantaneous release of pure energy employed in previous Primakoff models of \citet{o'brien92}. The structures and shock evolution are described and compared to that of previous models. An updated radiative cooling implementation is reported and an exploration of the simulations' parameter space is carried out in order to assess the impact of the various parameters on the results. We then go on to discuss further development of the model.

\section{Hydrodynamical simulations}\label{s2}

\subsection{The numerical scheme}\label{s3}

The hydrodynamical model for RS Ophiuchi was created using the \textsc{asphere} code (an updated version of the code used in \citealt{o'brien94}); an Eulerian one-dimensional (spherically symmetric) second order Godunov code \citep{godunov59}. It is based on the finite difference scheme of \citet{falle91} to solve the inviscid Euler equations of fluid flow in spherical polar coordinates which account for conservation of mass, momentum and energy. The internal energy density of the gas is
\begin{equation}
E = \frac{P}{\gamma - 1} + \frac{\rho u^{2}}{2} 
\end{equation}
where $P$, $\rho$ and $u$ are the gas pressure, density and velocity respectively, and $\gamma$ is the ratio of specific heats (= 5/3 for a monatomic gas). The temperature of the gas assumes an ideal gas equation of state for which 
\begin{equation}
T = \displaystyle \frac{\bar{m}}{k_{B}} \frac{P}{\rho}
\end{equation}
where $\bar{m}$ is the average particle mass for solar abundances and $k_{B}$ is Boltzmann's constant. The computational grid is divided into concentric radial spherical shells with steadily increasing widths as the radius increases. The stellar wind is produced using an appropriate in-flow boundary condition which is varied with time to obtain the desired evolution of wind density and speed. In the Primakoff models, the grid is initially filled with a $\rho \propto 1/r^{2}$ RG wind and energy is injected in a small region at the centre.

\subsection{Primakoff and adiabatic wind models}\label{s4}

In order to compare previous models based on the Primakoff similarity solution (both analytical and numerical Lagrangian, as in \citealt{o'brien92}) to our interacting winds model, we re-run the Primakoff simulations using \textsc{asphere}. A Primakoff situation consists of a spherically symmetric cool medium with density $\propto 1/r^{2}$ at the centre of which energy is injected at a point. The initial stationary distribution of density, $\rho$, is given by
\begin{equation}
\rho = \displaystyle \frac{\dot{M}}{4\pi r^{2} u}
\end{equation}
where we take the ratio of mass-loss rate into the RG wind $\dot{M}$ and its velocity $u$ to be $\dot{M}/u = 6.0\times 10^{12} \mathrm{g}~\mathrm{cm}^{-1}$ \citep{o'brien92}. An energy $E_{0} = 8.62\times10^{43}$erg is evenly distributed over a central region of radius three cells, corresponding to about $1.5\times 10^{11}$cm (the value for $E_{0}$ was chosen to match that of the wind model described hereafter for which the parameters were derived from various observations, see later). In a Primakoff problem, no energy loss due to radiation is included. The analytical solution \citep{chevalier82} for the shock radius as a function of time $t$ is
\begin{equation}
r_{s} = a t^{2/3}
\end{equation}
where 
\begin{equation}
a = \left( \displaystyle \frac{6 E_{0}}{\dot{M}/u}\right)^{1/3}~~,
\end{equation}
whilst the density, pressure, velocity and temperature distributions are well defined functions of radius and time.

\begin{figure*}
\begin{center}
\includegraphics[scale=0.33]{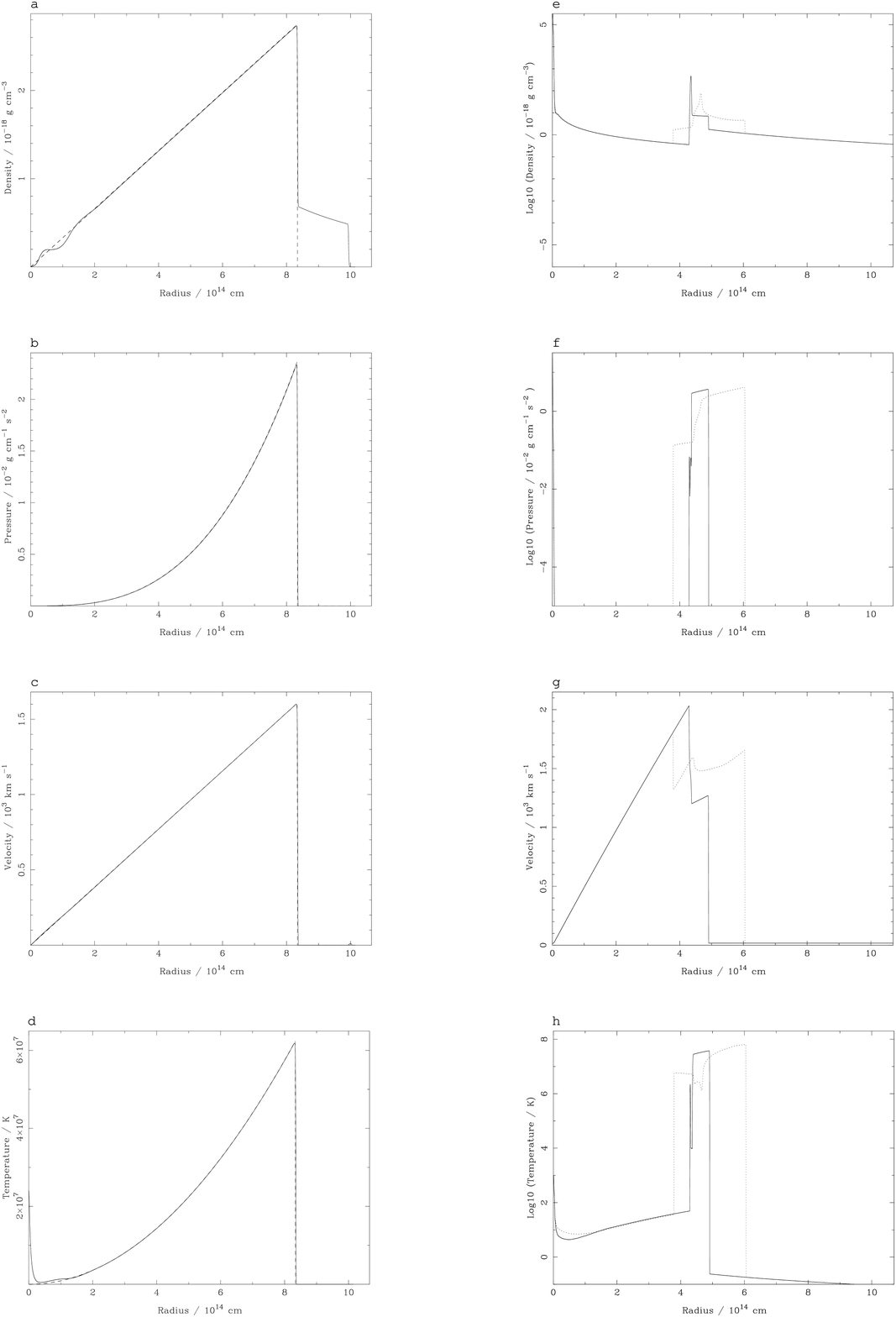}
\caption{Left: gas density $\rho$ (a), pressure $P$ (b), velocity $v$ (c) and temperature $T$ (d) as a function of radius calculated by \textsc{asphere} (using 6000 cells) without radiative cooling (solid line) for the Primakoff simulation compared to the adiabatic analytical solutions (dashed line) at day 30 after outburst. The outburst energy is $8.62\times 10^{43}$erg. The other parameter values are given in the text.\protect \\ Right: $\log(\rho)$ (e), $\log(P)$ (f), velocity (g) and $\log(T)$ (h) as a function of radius for the wind model (16000 cells) with the same outburst energy at day 30. The dotted line represents the adiabatic solution (section \ref{s4}) and the solid line is the cooled simulation (see sections \ref{s5} and \ref{s6})}
\label{1}
\end{center}
\end{figure*}

Fig.~\ref{1} (a to d) shows the \textsc{asphere} results for the different quantities along with the analytical solutions at a nominal time of 30 days after outburst. A forward shock is clearly visible at a radius of about $8.3\times10^{14}$cm. The matter to the right of the shock is the unshocked RG wind in which the $1/r^{2}$ dependence is clearly visible.

Previous models of RS Ophiuchi, as in \citet{o'brien92}, used a Primakoff-type solution with point-injection of energy at the centre of a RG wind to describe the dynamics. As the first X-ray observations were only performed on day 55 after outburst when the remnant was believed to already be in Phase II of its evolution (where a hot adiabatic blast wave is driven into the RG wind), it was decided that the manner in which the energy was injected at early times was unimportant and that a Primakoff-type solution would be adequately suited to the problem. In 2006, with X-ray observations starting at outburst and with the ongoing detailed monitoring of the system, it is necessary to think more deeply about the physics of the outburst and consider all three phases of the remnant evolution. A more realistic way of modelling the system is to treat the outburst as a sudden increase in wind velocity and mass-loss rate from the central WD.

A small region (20 cells) in the centre is used as the core from which material is injected into the grid. The RG wind is initially blown for 21 years (corresponding to the time between the 1985 and 2006 outbursts), filling the circumstellar region. The velocity of the RG wind is assumed to be $V_{1} = 15$\kms and its mass-loss rate $\dot{M}_{1} = 9.0\times 10^{18}~\mathrm{g}~\mathrm{s}^{-1}$ (see \citealt{o'brien92}). These parameters are then linearly increased to the outburst values $V_{2} = 3000$\kms (consistent with very high optical line width velocities observed by \citealt{buil06}, infrared line widths in \citealt{evans07} and Paper I X-ray temperatures) and $\dot{M}_{2} = 4.607\times 10^{21}~\mathrm{g}~\mathrm{s}^{-1}$ (see below) over a period of one day (an instantaneous increase to the outburst values causes the code to fail due to an overly large discontinuity) and are then kept on a constant plateau for another 4 days. Finally, they are linearly decreased back to the original values appropriate to the RG wind over 2 days, giving a total ejected mass of $M_{\mathrm{ej}} = 1.1\times 10^{-6} \msun$ and injected energy $E_{0} = 8.62\times 10^{43}$erg, for a total ejection phase duration of 7 days. These timescales were chosen to agree with the X-ray results of Paper I where the velocities in the system are observed to start decreasing after about 6 days. They are also in agreement with the TNR models of \citet{yaron05} who observe mass-loss phases lasting about 5 days in the outbursts of high-mass WDs. The total ejected mass was chosen to match the best estimate from \citet{o'brien92}, which is also consistent with \citet{hachisu07} who found $M_{\mathrm{ej}} \sim 2-3\times 10^{-6} \msun$.

The results obtained 30 days after outburst are plotted in the right column of Fig.~\ref{1} (e to h, dotted line) where density, pressure and temperature are plotted on a logarithmic scale for extra detail. The density plot reveals the presence of a strong forward shock around $6.05\times10^{14}$cm followed by a high-density contact discontinuity at $4.70\times10^{14}$cm and a reverse shock at $3.80\times10^{14}$cm. The velocity as a function of radius is linear up to the reverse shock. The contact discontinuity has somewhat lower temperature than the rest of the shell and a low velocity, along with the reverse shock. In comparison to the Primakoff solution, we note that the forward shock has not travelled as far due to the extended period over which the energy is injected. The other obvious difference is the presence of a reverse shock and a contact discontinuity arising from the fast wind $-$ slow wind interaction. These two extra components inside the hot shell are likely to affect the evolution of the ejecta, especially if internal energy or pressure, which is driving the shock forward, is lost through radiation.

\subsection{Radiative cooling}\label{s5}
Energy losses via radiative cooling can significantly affect the dynamics of a system. The cooling rate $\Lambda (T)$ as a function of gas temperature for a plasma of typical abundances was taken from \citet{raymond76}. We note that below $10^{4}$K, radiative cooling becomes very ineffective. Above $10^{8}$K, all the medium will be ionised and the gas will only radiate via free-free (Bremsstrahlung) emission which can be described by a simple $\Lambda (T) \propto T^{1/2}$ cooling law. $\Lambda (T)$ was tabulated and at each timestep, we subtract an amount of energy from each cell corresponding to an interpolation from the full cooling curve.

When incorporating radiative cooling into a numerical scheme another constraint needs to be taken into account, as the minimum timestep is not only limited by the \citet{courant67} condition (the dynamical timescale) but also by a cooling timescale. The cooling timescale is an approximation of the time it would take for the cell to lose all of its energy. Both dynamical and cooling times are computed for each cell, and the timestep is taken to be the shorter of the dynamical time and 5\% of the cooling time.

To confirm the validity of this method for radiative cooling we used once again the Primakoff similarity solution as a test-bed. A new run was performed, this time including radiative losses, and the results obtained were consistent with the linearised solutions derived in \citet{o'brien87} where cooling was treated as a first order perturbation. As opposed to previous cooling law approximations, we now account for the full range of temperatures seen in the simulations, including the important hydrogen peak around $1.7\times10^{4}$K.

\subsection{The cooled wind-driven shock model}\label{s6}

The system of forward and reverse shocks in RS Ophiuchi is expected to radiate strongly in the X-ray and energy losses are likely to be significant. The same wind simulation as in section 2.2 was performed but this time including cooling effects from time $t = 0$. The results are displayed on Fig.~\ref{1} (e to h, solid line). We observe the density of the contact discontinuity to be higher than in the adiabatic solution by a factor of $\sim 10$. The thickness of the shell is smaller and the forward shock has not travelled as far (the forward shock radius has decreased by a factor of $\sim 1.24$). The presence of the reverse shock is much less obvious as it is much closer to the contact discontinuity. It is also well cooled, along with the contact discontinuity. In the adiabatic case, we notice that the contact discontinuity is at a temperature of about $10^{6}$K which is close to the peak cooling rate, thus consistent with the fact that it appears to be strongly radiating in the cooled run.

\begin{table}
\begin{center}
\caption[Fractions of the total radiated energy]{Fractions of the total radiated energy\\}
\begin{tabular}{@{}lccr@{}}
\hline
\hline
Time & Shocked & Shocked & Other\tablenotemark{*}\\
(days) & ejecta & RG wind &    \\
\hline
1 & 24.5 \% & 18.6 \% & 56.9 \%\\
5 & 40.3 \% & 50.9 \% & 8.8 \%\\
10 & 32.4 \% & 63.3 \% & 4.3 \%\\
50 & 25.2 \% & 72.6 \% & 2.2 \%\\
100 & 23.1 \% & 75.0 \% & 1.9 \%\\
\hline
\end{tabular}
\tablenotetext{*}{\protect Largely within contact discontinuity}
\tablecomments{Time is in days after outburst. See text for a definition of the different parts of the system.}
\label{t1}
\end{center}
\end{table}

Table~\ref{t1} lists the fractions of the total energy-loss via radiative cooling from the different parts of the shell at five different epochs. The shocked ejecta is defined as the region between the reverse shock and the contact discontinuity, and the shocked RG wind is situated between the contact discontinuity and the forward shock. We observe the shocked ejecta to radiate a large portion of the energy at very early times. As a result, it tends to disappear relatively quickly (as seen on Fig.~\ref{1}) with the reverse shock virtually reaching the contact discontinuity as the energy that was driving the shock away from the contact discontinuity is lost. This is a major difference with previous Primakoff models as they did not account for the presence of both reverse shock and shocked ejecta. Once the shocked ejecta has cooled, the part of the shell which starts to dominate the energy-loss is the shocked RG wind which has radiated 3/4 of the total radiated energy 100 days after outburst. However, even at late times the shocked ejecta remains an important contributor to the total radiated energy, thus showing that the reverse shock plays a major role in the cooled dynamics of the system, greatly affecting its evolution. Finally, the thin, high density contact discontinuity is inevitably spread over a small number of cells in our numerical model and these high density cells are expected to contribute to the cooling in non-negligible proportions, since the cooling rate scales as $\rho^{2}$. However they do not dominate the energy losses over the shocked shells. Virtually all of the radiated energy, over all but the first day or so, is emitted by the shell of shocked material.

\begin{figure}
\begin{center}
\includegraphics[scale=0.48]{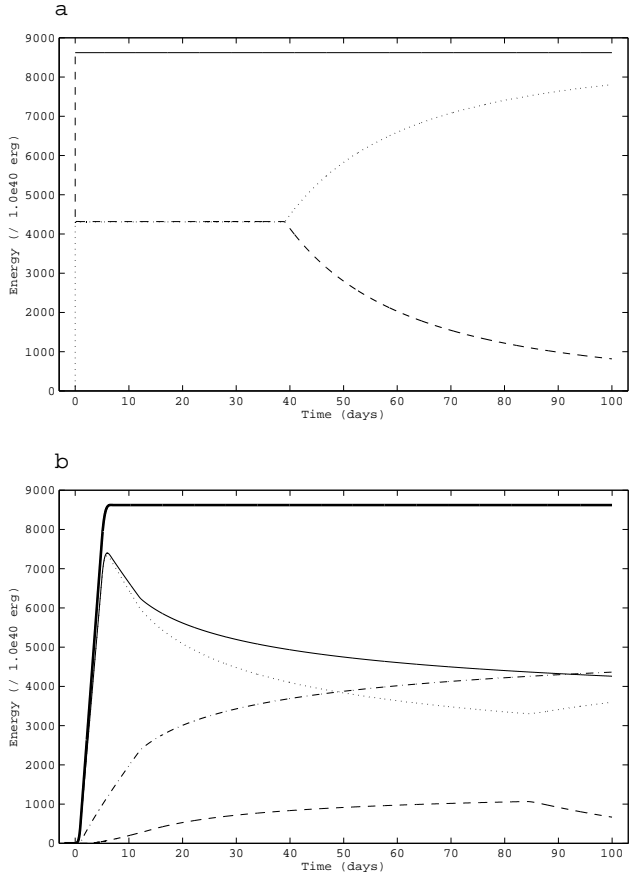}
\caption{Energies as a function of time. (a) Primakoff: the solid line represents the total energy, the dashed line is the thermal energy and the dotted line is the kinetic energy. (b) Wind model: same as for (a) plus dot-dash line is the cumulative radiated energy and the bold solid line is the sum of the radiated and total energies. The change in behaviour at $\sim40$ days in (a) and $\sim84$ days in (b) is due to the shock break-out, see text.}
\label{2}
\end{center}
\end{figure}

To check the consistency of our simulations we plot in Fig.\ref{2} the total, thermal and kinetic energies in the grid as a function of time. The results for the adiabatic Primakoff and cooled wind model are displayed. In addition, for the cooled case we have included the cumulative radiated energy and the sum of the total and radiated energies.

We see that in the adiabatic Primakoff case the total energy in the grid remains constant, as expected, and the energy is conserved up to an accuracy of 0.0015\%. The energy is initially all in thermal form but is rapidly shared evenly between thermal and kinetic. We observe the forward shock to break out of the RG wind at $t\approx39$ days where the material expands out into the low-density ISM, leading to much higher kinetic energy and adiabatic cooling hence the bifurcation of thermal and kinetic energy at this time.

In the case of the cooled wind model, the total remnant energy increases from 0 to 7 days (duration of the fast wind phase) and subsequently decreases due to energy loss via radiation. Radiated energy is seen to increase rapidly at early times and slows down from about 11 days onwards, supposedly once the material is relatively well cooled. The fact that most of the radiated energy is lost during the first 10 days explains why the shocked ejecta retains a large proportion of the radiated energy through the run, even though it is not clearly visible in the shell at later times. $E_{0}$ is injected mainly in the form of kinetic energy (fast wind) and we observe the break-out of the RG to occur much later than in the Primakoff case ($\sim84$ days). We finally note that energy is conserved in the system as total and radiated energies add up to a constant value equal to the outburst energy, to an accuracy of 0.018\%.

\begin{figure}
\begin{center}
\includegraphics[scale=0.5]{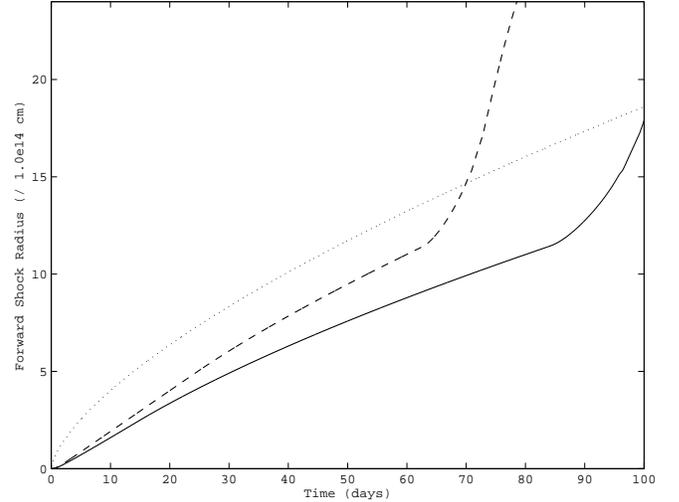}
\caption{Position of the forward shock as a function of time. The dotted line represents the analytical adiabatic $a t^{2/3}$ Primakoff solution for $E_{0} = 8.62\times 10^{43}$erg, the dashed line is the adiabatic wind model and the solid line is the cooled wind model.}
\label{3}
\end{center}
\end{figure}

Fig.~\ref{3} shows the position of the forward shock as a function of time, compared to that of an analytical Primakoff solution. We observe the Primakoff shock to expand very rapidly at early times. The wind model displays a more linear expansion, in agreement with the initial VLBI radio observations of the expanding shock wave carried out by \citet{o'brien06}, although they only have three early measurements of the shock radius and the Primakoff evolution cannot be completely rejected. As expected, the cooled model sees the forward shock expanding more slowly. The fact that the wind model also includes the ejection of mass as opposed to energy alone in the Primakoff case can explain why the shell seems to have more inertia at early times thus taking longer to accelerate, and also why the shell appears to be carrying more momentum hence decelerating later on.
 
The kinks in the wind model curves correspond to the time when the forward shock breaks out of the surrounding RG wind. The effect of this phase of evolution on the X-ray emission of RS Ophiuchi was explored in detail by \citet{o'brien92} who claimed it could account for some of the flux decrease seen 62 days after outburst. It is now believed that during this phase of evolution the X-ray flux is in fact dominated by the WD which is undergoing a SSS phase (\citealt{osborne06a}; \citealt{hachisu07}; Osborne et al. in prep.) due to ongoing nuclear burning on its surface. Once the SSS phase is over \citep{osborne06b}, the emission from the shock is again dominant.

\begin{figure}
\begin{center}
\includegraphics[scale=0.5]{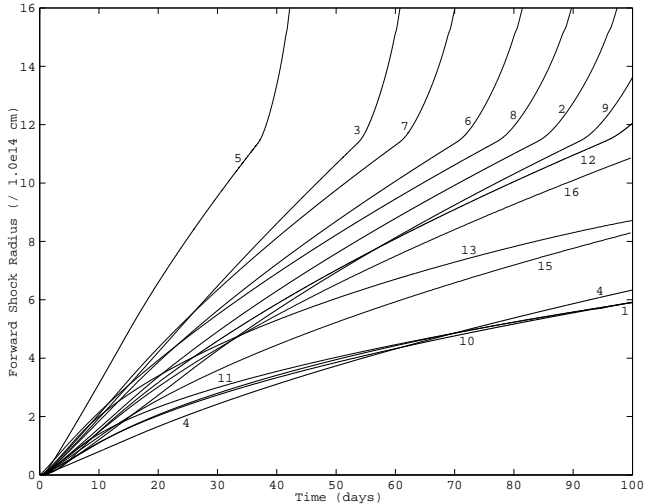}
\caption{Position of the forward shock as a function of time for runs 1 to 16 (run 14 excluded).}
\label{4}
\end{center}
\end{figure}

We now vary the three most important parameters in our wind model: the injected energy $E_{0}$, the ejected mass $M_{\mathrm{ej}}$ and the fast wind phase duration $t_{0}$. Table~\ref{t2} lists the different parameters employed in the various runs, and the results are displayed on Fig.~\ref{4}.

\begin{table*}
\begin{center}
\caption[Model parameters for Runs 1 to 16]{Model parameters for Runs 1 to 16\\}
\begin{tabular}{@{}ccccccc@{}}
\hline
\hline
   Run & Slow wind & \multicolumn{2}{c}{Fast wind} & Fast wind & Outburst & Ejected\\
       & mass-loss & mass-loss & velocity & phase duration & energy & mass\\
       & $\dot{M}_1$ & $\dot{M}_2$ & $V_{2}$ & $t_{0}$ & $E_{0}$ & $M_{\mathrm{ej}}$\\
       & $(\times 10^{18}~\mathrm{g}/\mathrm{s})$ & $(\times 10^{18}~\mathrm{g}/\mathrm{s})$ & (km/s) & (days) & $(\times 10^{43}$erg) & $(\times 10^{-6} \msun)$\\
   \hline
1 & 9 & 1000 & 3000 & 7 & 1.85 & 0.24\\
2 & 9 & 4607 & 3000 & 7 & 8.62 & 1.10\\
3 & 9 & 20000 & 3000 & 7 & 36.94 & 4.78\\
4 & 9 & 4607 & 1398 & 7 & 1.85 & 1.10\\
5 & 9 & 4607 & 6251 & 7 & 36.94 & 1.10\\
6 & 9 & 4607 & 3500 & 7 & 11.58 & 1.10\\
7 & 9 & 4607 & 4000 & 7 & 15.13 & 1.10\\
8 & 9 & 32245 & 3000 & 1 & 8.62 & 1.10\\
9 & 9 & 2150 & 3000 & 15 & 8.62 & 1.10\\
10 & 40 & 4607 & 3000 & 7 & 8.62 & 1.10\\
11 & 40 & 500 & 12000 & 7 & 14.77 & 0.12\\
12 & 9 & 7239 & 3969 & 7 & 8.62 & 1.10\\
13 & 60 & 1000 & 16300 & 6 & 46.73 & 0.20\\
14 & 9 & 4607 & 3000 & N/A & N/A & N/A\\
15 & 9 & 1883 & 3000 & 7 & 3.48 & 0.45\\
16 & 9 & 3138 & 3000 & 7 & 5.80 & 0.75\\
\hline
\end{tabular}
\tablecomments{Run 2 is equivalent to that described in section \ref{s6}. The slow wind velocity is $V_{1} = 15\kms$ for all runs. Where $t_{0} \neq 7$, the relative proportions in time of rise, plateau and decline of mass-loss rate and velocity have been kept constant.}
\label{t2}
\end{center}
\end{table*}

There are two ways of varying the outburst energy, as both the fast wind velocity and mass-loss rate can be altered. In the latter case (runs 1, 3, 11 and 13) the ejected mass is affected. The deceleration rate of the forward shock appears to be dependent on the ejected mass rather than the outburst energy, as the shock in run 1 is observed to decelerate faster than that of run 4, and in the opposite sense for runs 3 and 5. We can see that the forward shock in run 1 is ahead of run 4 at early times, but is eventually caught up then overtaken by the run 4 shock. The difference in ejected mass means the shell carries more momentum; its initial acceleration is smaller but the greater momentum drives it forward for a longer period of time. We also see that increasing the density of the RG wind (run 10) causes stronger deceleration of the forward shock, as expected.

Keeping the mass-loss rate constant (runs 4 and 5) leads to extreme values for the velocity. Instead we have chosen to vary the fast wind velocity by small amounts (runs 6 and 7) keeping $\dot{M}_2$ constant in order to assess the true role of the ejecta velocity. We see that although the increase in injected energy is only of a few $\times 10^{43}$erg compared to run 2, the ejecta speed has an important effect on the shock evolution as predictions of the time of RG wind break-out differ by up to $\sim25$ days in the case of run 7.

Finally, we see the fast wind phase duration not to have a major effect on the simulations when $E_{0}$ and $M_{\mathrm{ej}}$ are kept constant. The forward shocks in runs 8 and 9 remain fairly close to run 2 throughout the 100 days, and the difference in RG break-out time is about 10 days between the two extremes in fast wind phase duration. We do note that as $t_{0}$ tends to 0 (run 8), the profile of the shock radius as a function of time increasingly resembles that of a Primakoff model (fast expansion at early times), which makes sense in terms of the energy being released almost instantly in the system. However, the forward shock still goes through a short slow expansion phase just after the start of the outburst.

An interesting additional result is that the reverse shock is visible in runs 1, 5 and 9, suggesting that high velocity, low-mass and lengthy ejections favour the detachment of the reverse shock from the contact discontinuity. The reverse shock is present in all the runs but in the other cases it has almost merged with the contact discontinuity. In runs 1, 5 and 9 it is well behind the contact discontinuity, with run 5 being the most evident case.

\section{Forward shock velocities and early {\it Swift} data}\label{s7}

\begin{figure}
\begin{center}
\includegraphics[scale=0.48]{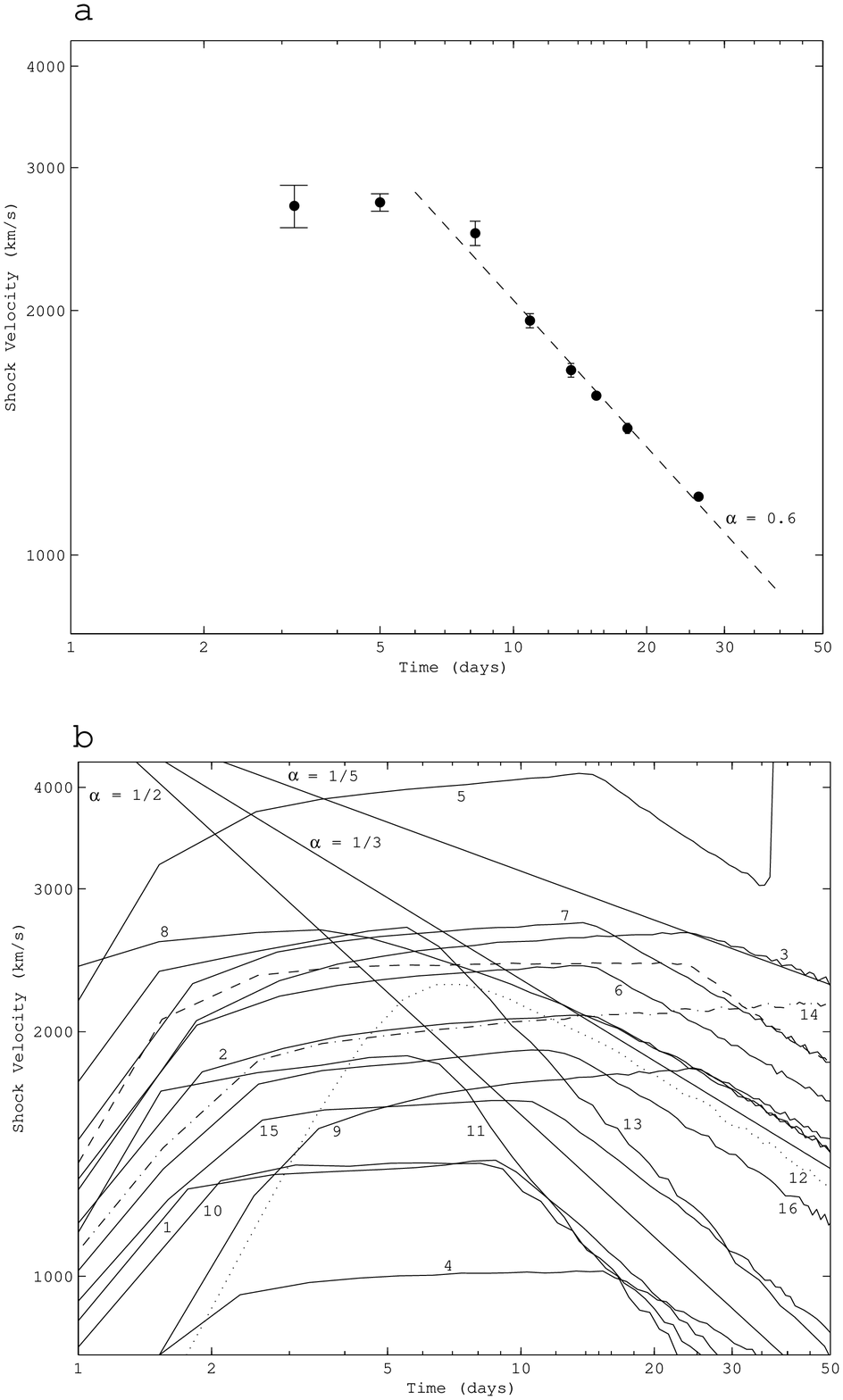}
\caption{(a) The Paper I (Fig.~4) early shock velocities as a function of time obtained from single temperature fits to the X-ray data. The shock velocity appears to follow the power law $v_{s} \propto t^{-\alpha}$ with $\alpha = 0.6$.\protect\\
(b) Forward shock velocity as a function of time for runs 1 to 16 (see Table~\ref{t2} for a list of parameters). The solid straight lines represent analytical predictions for Phase I of the remnant evolution with $\alpha = 1/5$, Phase II where $\alpha = 1/3$ and Phase III where $\alpha = 1/2$. The kink in the run 5 curve circa 37 days corresponds to the RG wind break-out of the shock. See text for a full description.}
\label{5}
\end{center}
\end{figure}

The ultimate aim of our simulations is to model the X-ray emission from the shocks in the interacting winds of RS Ophiuchi. In Paper I, single temperature emission models were fitted to the X-ray data and shock velocities as a function of time were derived, as reproduced on Fig.~\ref{5}a. The results were broadly in agreement with models in \citet{bode85} which predicted Phase I to be over after $\sim 6$ days. However, the shock evolution was observed to follow $v_{s} \propto t^{-\alpha}$ with $\alpha = 0.6$, a deceleration rate greater than theoretical expectations, even for a remnant in Phase III where $\alpha = 1/2$. The data also reveal a sharp turnover from a seemingly constant velocity stage to a deceleration stage, rather than a smooth transition between phases of evolution.

The results from our wind models are displayed on Fig.~\ref{5}b (see Table~\ref{t2} for the list of parameters). The plots are compared to analytical predictions of the remnant evolution where the forward shock velocity $v_{s} \propto t^{-\alpha}$ with $\alpha = 1/5$ in Phase I, $\alpha = 1/3$ in Phase II, and $\alpha = 1/2$ for Phase III. We show that while velocity gradients are fixed for the analytical solutions, different gradients in velocity are obtainable in our model simply by altering wind parameters. All the curves on Fig.~\ref{5}b present the same trend. They are composed of an early increase in forward shock velocities, a slowly rising plateau (apart from run 12 which doesn't have a plateau), a sharp turning point and a deceleration phase.

The initial acceleration phase is related to the increase in velocity and mass-loss rate from the slow wind to fast wind states, as it is over after 2 days in almost all runs, apart from run 9 for which the transition between slow and fast wind takes twice as long.

The plateau in shock velocity has two defining traits: its magnitude and duration. By looking at runs 2, 4, 5, 6 and 7, we can see that its height is defined by the ejection velocity of the fast wind $V_{2}$, as all the curves present the same shape, simply being shifted up and down the velocity axis. However, this is not the only defining parameter, as illustrated by run 10 for which the fast wind velocity is the same as that of run 2, but the slow wind density is higher. The forward shock is running into a denser medium, and consequently its velocity is greatly decreased.

The end of the fast velocity plateau is marked by a sharp turnover. It was thought in \citet{bode06} that this marked the end of Phase I (free expansion) of the remnant evolution. A first order approximation for the time at which this occurs is when the mass swept-up by the forward shock $M_{\mathrm{swept}}$ is equal to the ejected mass $M_{\mathrm{ej}}$. By considering runs 2, 4, 5, 6 and 7, it is evident that this cannot be the case here. Indeed, all these runs have the same $M_{\mathrm{ej}}$ and a turnover occuring virtually at the same time. If the fast wind velocity in run 5 is much higher than in run 2, its forward shock travels outwards much faster (as seen on Fig.~\ref{4}) and it will thus have swept-up the required mass at a much earlier time, leading to a much earlier turnover. Run 12 (dotted curve) has the same outburst energy and ejected mass as run 2, but there is no plateau in $\dot{M}$ and $V_{2}$ during the outburst. Instead it has a simple linear rise and fall in $\dot{M}$ and $V_{2}$ equally distributed over the duration of the fast wind phase. It does not show a plateau in shock velocity, simply an increase then decrease, with the rate of the latter being the same as run 2. In run 14 (dot-dash), the same slow and fast wind mass-loss rates and velocities as in run 2 are employed, but the fast wind is not switched off. We can see that no turnover is observed. We can thus conclude that the plateau in forward shock velocity is a result of the plateau in $\dot{M}$ and $V_{2}$ during the fast wind phase of the outburst.

The turnover is in fact directly related to the end of the fast wind phase; the turnover occurs when the last of the fast wind material ejected at the end of the plateau reaches the shocked shell, after which there is no additional energy input to help drive the shock. The sharpness of the turnover is in agreement with Fig.~\ref{5}a.

The manner in which the mass is ejected controls the evolution of the shock. However, a transition from Phase I to Phase II of remnant evolution is still expected to be observed. This is best visible in the case of run 8 where the fast wind phase is very short and should not be affecting the subsequent evolution of the remnant. A phase of shock deceleration with a gradient $\alpha = 1/5$ is visible in the range $5\lesssim t \lesssim 20$ days after which the forward shock is observed to adopt a Phase II type behaviour with $\alpha = 1/3$. The transition between the two phases is very smooth (much more so than the turnover after velocity plateaus seen in the other runs), which is expected for a transition between two states of evolution which are valid in different limits, and thus does not fit the observations.

The final aspect of the curves to be addressed is the deceleration rate of the forward shock after the turnover. In all cases, this appears to have a constant value. There appear to be more possible values for $\alpha$ than just the three analytical limits. Firstly, we note that the high $M_{\mathrm{ej}}$ run 3 is in agreement with the Phase I gradient for which the swept-up mass is considered not to have an effect on the remnant evolution. We shall call this a class 1 run. Class 2 runs (runs 2, 4, 5, 6, 7, 8, 9 and 12) display a Phase II behaviour, the limit in which cooling effects are treated as negligible. Indeed, the adiabatic run 2 (dashed curve) displays an $\alpha = 1/3$ gradient, and an adiabatic version of run 13 (not plotted here) also showed the same behaviour. We can thus safely conclude that $\alpha$ cannot be greater than 1/3 if radiative cooling is not included. We can also confirm that cooling is not significant in the runs listed above.

Class 3 runs (runs 1, 10, 11 and 13) show $\alpha \ge 1/2$ which is not predicted by analytical models. The Phase III approximation ($\alpha = 1/2$) is used to describe the evolution of a well cooled remnant, i.e. where radiative losses are significant. The fractions of the blastwave energy which is radiated away 50 and 100 days after outburst are listed in Table~\ref{t3}. We note that the blastwave in run 3 radiates only a small fraction of its energy. The Phase II type runs appear to radiate around half their energy, which suggests that this still does not classify as `significant' energy loss. Finally, the high-$\alpha$ runs radiate a large fraction of their energy (circa $80\% - 90\%$). The gradients of runs 15 and 16 lie between classes 2 and 3, with run 16 being very marginally steeper than $\alpha = 1/2$. This represents the limit at which cooling starts to affect the evolution of the shock, ie $E_{\mathrm{rad}}/E_{0} \sim 0.6$.

\begin{table}
\begin{center}
\caption{Fractions of the total energy which is radiated away\\}
\begin{tabular}{@{}ccc@{}}
\hline
\hline
Run & \multicolumn{2}{c}{$E_{\mathrm{rad}}/E_{0}$ (\%)}\\
    & Day 50 & Day 100\\
\hline
1 & 77.3 & 83.9\\
2 & 44.9 & 50.6\\
3 & 21.2 & 24.5\\
4 & 46.9 & 59.2\\
5 & 31.8 & 34.5\\
6 & 43.5 & 48.1\\
7 & 42.1 & 45.8\\
8 & 40.1 & 45.3\\
9 & 47.9 & 54.4\\
10 & 79.5 & 85.2\\
11 & 91.5 & 93.9\\
12 & 46.1 & 51.5\\
13 & 87.3 & 91.4\\
15 & 68.1 & 75.6\\
16 & 56.3 & 63.1\\
\hline
\end{tabular}
\tablecomments{Time is in days after outburst. $E_{\mathrm{rad}}$ is the radiated energy. Run 14 is excluded as energy is continuously injected in the grid and the fraction that the radiated energy represents becomes meaningless.}
\label{t3}
\end{center}
\end{table}

By considering, runs 16, 15, 1, 10, 13 and 11, a correlation is visible between the steepness of the shock velocity gradient and the amount of energy radiated: the higher the fraction, the higher the $\alpha$. For these runs, the fraction of energy radiated appears to depend on the ratio of ejected mass to RG wind density, as ratios decrease from runs 16 to 11.

In order to obtain a steep gradient as in Fig.~\ref{5}a and keep the RG wind density reasonably low, a low ejected mass is required (in agreement with Osborne et al. in prep.). However, if high shock velocities suggested by the Paper I data are to be attained, a very (perhaps unrealistically?) high fast wind velocity is needed, as illustrated by run 13, and this is always true for a low $M_{\mathrm{ej}}$ to RG wind density ratio.

An attempt to perform a detailed fit to the Paper I data does not seem appropriate here as the multi-temperature structure inside the hot shell revealed in our models shows that fitting single temperatures to the system is likely to be inadequate. In addition, we have assumed spherical symmetry which is almost certainly not appropriate (see next section). In a following paper, we will calculate the predicted X-ray spectra directly from our hydrodynamical simulations using many temperature components and compare the results to the data. In order to model the X-ray spectra beyond $t\sim 26$ days, we will also have to substract the SSS component from the emission.

\section{Conclusions}\label{s8}

By describing the mass ejection in the form of a wind, we have presented a more realistic approach to modelling the dynamics of RS Ophiuchi. The fast/slow wind interaction leads to the formation of a forward shock, a reverse shock and a contact discontinuity. The hot shell, sandwiched between the two shocks, travels outwards sweeping up the RG wind and eventually breaking out of it, freely expanding into the low-density ISM. Outburst parameters $E_{0}$ and $M_{\mathrm{ej}}$ are intrinsically linked through the fast wind mass-loss rate $\dot{M}_2$ and a non-zero fast wind phase duration is less artificial than an instantaneous point-injection of energy. The forward shock is consistently seen to undergo near linear expansion at early times, independent of parameter values, with the exception of an extremely short fast wind phase duration and very small ejecta mass. The cooling power-law approximation has been replaced by a high-accuracy cooling curve for increased realism. Our models are successful in reproducing the general traits of the Paper I shock velocities: rise to a plateau terminated by a sharp turnover followed by a power-law deceleration. For well cooled remnants, deceleration rates of $\alpha > 1/2$ are achievable for low mass ejecta, and they are seen to correlate with the amount of energy radiated away. However, the models were not able to satisfactorily reproduce the high shock velocities observed without requiring extremely high fast wind velocities.

\citet{o'brien92} introduced radiative cooling only 10 days after outburst as the high cooling rates in the early stages of the simulation were forcing the timesteps to be very small, leading to impossibly long run times. The primordial behaviour of the ejecta was believed not to affect the late evolution of the remnant. With modern computers, we can now run the simulations with the cooling switched on from the start and this appears to greatly affect the dynamics even 100 days after outburst. Indeed, as seen in Fig.~\ref{2}b, most of the radiated energy is lost during the first 10 days. Simulations were run with varying cooling delays and the position of the forward shock as a function of time was significantly altered. Delaying the cooling for 10 days resulted in advancing the time of RG shock break-out by approximately 35 days.

A low ejected mass has very important consequences on the long term evolution of RS Ophiuchi. If it is smaller than the mass accreted between two consecutive outbursts, the WD must then be growing in mass, ultimately exploding as a supernova once it reaches the Chandrasekhar limit (e.g. Paper I).

Finally, following the radio observations of \citet{porcas87}, \citet{taylor89} and \citet{o'brien06}, the theoretical work of \citet{lloyd93} and the HST imagery of \citet{bode07} on day 155, it is now believed that the nebular remnant of RS Ophiuchi has a bipolar structure. The models will therefore need to be extended to at least two dimensions in which case the break-out of the RG wind will no longer be a singular event in time but could be spread out over many days. In addition, the very thin cooled contact discontinuity will be subject to Rayleigh-Taylor and thin-shell instabilities which can only be properly modelled in three dimensions. Extension of our models to three dimensions is being pursued.

\acknowledgments

NMHV is supported by a University of Manchester research studentship. MFB acknowledges the support of a PPARC Senior Fellowship. We would like to thank the referee for useful comments which led us to a deeper understanding of the models and greatly improved this paper as a result.

\clearpage

\end{document}